\documentclass[12pt]{article}
\usepackage{blois,epsfig}

\begin{document}

\vspace{4mm}
\title{Baryon Transverse Momentum Distributions at LHC and the "Knee" in Cosmic Proton Spectrum}

\author{O.I.PISKOUNOVA}

\footnote{O.I.PISKOUNOVA, e-mail: piskoun@sci.lebedev.ru, tel. +7(499)1326317, P.N.Lebedev Physics Institute of Russian Academy of Science, Leninski pr. 53, 119991 Moscow, Russia}

\address{P.N.Lebedev Physics Institute of Russian Academy of Science, Moscow, Russia}

\maketitle\abstracts{ The transverse momentum spectra of $\Lambda^0$ hyperon from
LHC experiments (ALICE, ATLAS, CMS) are considered from the point of view of Quark-Gluon String Model (QGSM) 
as examples of typical baryon spectra at very high energies. The LHC data at 
$\sqrt{s}$ = 0.9 and 7 TeV and STAR data at $\sqrt{s}$ = 200 GeV are fitted with the universal QGSM formula that includes the energy dependent slope as the main parameter. The dependence of <$p_t$> on $\sqrt{s}$ has been obtained with the help of this formula. The asymptotics of the energy dependence of average transverse momenta shows the expected pomeronic behavior $ ~ s^ \Delta_P(0)$, where $\Delta_P(0) = \alpha_P(0)-1 $= 0.12 and $\alpha_P(0)$ is the intercept of Pomeron regge trajectory that was estimated in the previous QGSM studies. This conclusion is very important for cosmic ray physics. It means that the long debated "knee" in the cosmic proton spectra at $E_p= 4 10^15$ eV in laboratory system can not be considered any more as the result of dramatic changes in the dynamics of baryon hadroproduction. In the same time it may indicate the maximal energy of proton that are beeing produced in our Galaxy. This last conclusion depends on the hadroproduction model.}

\section{Introduction}
The transverse momentum spectra of $\Lambda^0$ hyperons that were measured by many hadron collider experiments: STAR \cite{star}, UA1 \cite{ua1}, UA5 \cite{ua5}, CDF \cite{cdf} and experiments at LHC \cite{alice,atlas,cms} show the form of distributions, see Fig.1, which is changing from lower energy to higher. As it was established in the previous publication \cite{antiproton}, there is specific dependence of spectra on the type of colliding hadrons. Due to this difference, we have to confine ourself to the consideration of results of only proton-proton collision experiments : STAR, ALICE, ATLAS and CMS in the wide range of $\sqrt{s}$ energies: beginning from 200 GeV to  the up-to-date LHC energy 7 TeV. 

The suggested range of energy has motivated me to study the behaviors of baryon spectra, because the cosmic ray proton spectrum shows the "knee" at the very energy between Tevatron and LHC experiments \cite{knee}. The change in the slope of spectrum of protons, produced in space, may be of astrophysical origin otherwise it means a sustantial change in dynamics of particle production. We have to learn the behaviors of baryon production spectra from one energy to another in order to conclude that no dramatic changes have place in hadroproduction processes at "knee" energy and a little above.          

\section{Comparison of  transverse momentum spectra from LHC experiments}

The recent data on $\Lambda^0$ hyperon distributions are obtained in the following LHC
groups: ALICE \cite{alice} at 900 GeV, ATLAS \cite{atlas} at 900 GeV and 7 TeV and CMS \cite{cms} at 900 GeV and 7 TeV. The lower energy experiment, which the data of LHC groups can be compared with, is STAR \cite{star}($\sqrt{s}= 200 GeV$).  

In Fig.2 the essential flattering in the slope of $dN/dp_t$ can be seen, if we fit the data with a
simple exponential function $~ \exp{-B*p_t}$. We can conclude that transverse momentum spectra are
more and more hard with the energy growth, beginning from B=4,0 for STAR data ($\sqrt{s}$= 200 GeV) to B=1,8 at 900 GeV in ALICE. The slope is going to be more flat if we take the spectra at $\sqrt{s}$ = 7 TeV as it is seen in Fig.3, where B = 1,5. 

Both experiments, ATLAS and CMS, have presented the spectra with the similar slopes. The different form of the distributions at low $p_t$s may be caused by specifics in efficiency of the detecting procedure. It should be mentioned here that ATLAS spectra have systematically low efficiency at $p_t< 1$ GeV in comparison to the results of other LHC groups.

\section{Baryon transverse momentum distributions in QGSM}

It was studied in early QGSM paper \cite{veselov} that the transverse momentum spectra of hadrons after proton-proton collisions can be perfectly described with a bit more complicate dependence:

\begin{equation}
E \frac{d^{3}\sigma^H}{dx_F d^{2}p_{t}}= \frac{d\sigma^{H}}{dx_F}*A_0*\exp{-B_0*(m_t-m_0)},\nonumber 
\end{equation}
where $m_0$ is the mass of produced hadron, $m_t$ = $\sqrt{p_t^2+m_0^2}$ and $B_0$ used to bring the dependence on $x_F$, but in central region of rapidity this slope is constant. The slopes for
the spectra of many type of hadrons ($\pi$, K, p and $\Lambda^0$) were estimated for the data of proton-proton collisions of the energies available those times. The value of the slopes of baryon spectra was approximately $B_0$ = 6,0.  

Now we have to conclude that $B_0$ depends strongly on the energy . More, as it is seen from the spectra at LHC and RHIC, the value of $m_0$ is not the proton or hyperon mass.
The better description of hyperon spectra, what is shown in Fig.4, can be achieved  with $m_0$ = 0,5 GeV
that is kaon mass. Due to this fit the values of slope parameter should be equal for all LHC experiments of the same energies. It means that as soon as we estimate the slopes for certain collider
energy we obtaine a chance to build the dependence of average transverse momentum on the energy for high energy proton-proton collisions.

\section{Average transverse momenta at $\sqrt{s}$ = 200 GeV, 900 GeV and 7 TeV}

In Fig. 5 the resulting dependence of average $p_t$ on the energy of interaction is shown.
It is very predictable that the energy dependence of the average $p_t$ value is reproducing the typical multipomeronic behaviour of differencial production cross sections, which have been studied in QGSM withing three decades, see \cite{qgsm,hyperon,charm,beauty}. 
We could only suggest that the fast growing of $<p_t>$ before asympthotical regime is due to the production of  top-antitop quark pairs and other super heavy fundamental particles, which are unknown yet and  decaying finally into light baryons.
 
\section{Conclusions}

The review of results in transverse momentum distributions of hyperons that are produced in proton-proton collisions of various energies revealed a significant change in the slopes of spectra in the region of $p_t$ = 0,3 - 4,0 GeV/c. The spectra of baryons are becoming harder and harder with the energy growth from RHIC ($\sqrt{s}$=200 GeV) to LHC (0,9 and 7 TeV). The detailed analysis of hyperon spectra, which is analogous to our early study in frameworks of Quark-Gluon String Model, demonstrates the change of slopes from $B_0$ = 5,7 (STAR at 200 GeV) to $B_0$ = 2,1 (LHC at 7 TeV). 

As a result, the average $<p_t>$ value is growing very fast up to approximately $\sqrt{s}$= 1 TeV and then it goes with the asymptotics ~ $s^\Delta_P(0)$, where  $\Delta_P(0)$ = $1-\alpha_P(0)$ = 0,12 ($\alpha_P(0)$ is the intercept of pomeron regge trajectory, which is the main phenomenological parameter of the model). The asympthotics corresponds to typical multipomeron behavior of inclusive production cross sections. Thus it makes us conclude that processes taking place in baryon production at the up-to-date energies of LHC are not something unpredictable. This statement is very important for cosmic ray physics, where the "knee" (the change of the slope) at $E_{lab}$ $\approx$ 4* $10^{15}$ eV in cosmic proton spectra might origin in hadronic interactions. As we have discussed above, nothing unpredictable happens with baryon spectra up to $\sqrt{s}$ = 7 TeV corresponding to $E_{lab}= 2,5*10^{16}$ eV. It means that the "knee" is caused by astrophysical reasons. On my mind the "knee" may indicate the maximal energy of proton that are beeing produced in our Galaxy. But the idea of proton production in space assumes a futher detailed investigation of the dynamics of proton production by means of QGSM. The baryon/antibaryon asymmetries that were already measured in LHC experiments are intended to be discussed in our upcoming publication.
 
\section{Acknowledgments} 

I am thankful to Ludmila Malinina, who has found the ALICE group publication for me.

\newpage

\begin{figure}[h]
\begin{center}
\epsfig{figure=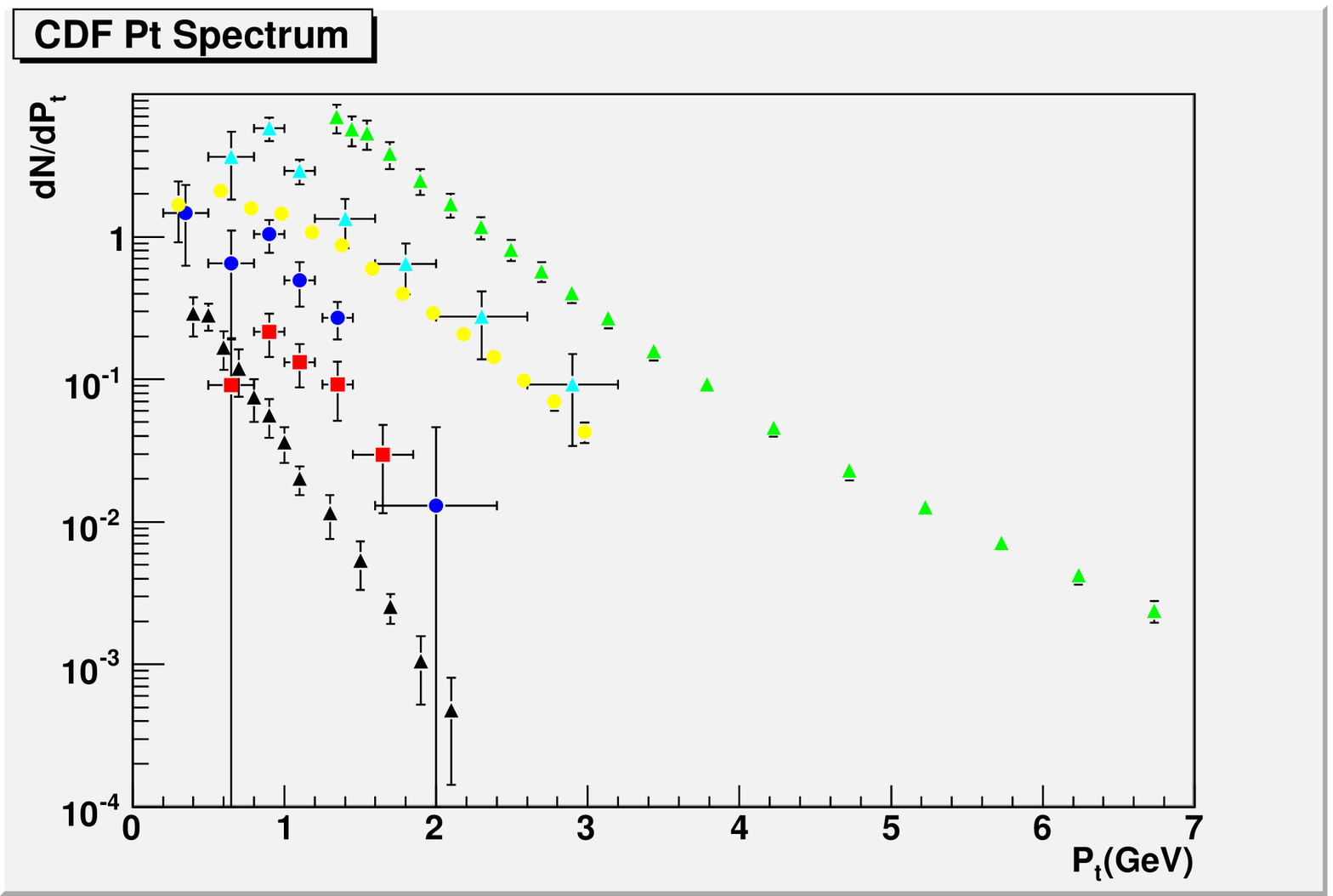,height=3.0in}
\caption{Transverse momentum distributions from different experiments and of various energies. The data are from STAR(200GeV)-black stars; UA5 energies: $\sqrt(s)$ = 200 GeV(red squares),
546 GeV(blue circles) and 900 GeV(aqua triangles); UA1(630GeV)-yellow circles and CDF(1.8 TeV)-green triangles.}
\end{center}
\end{figure}

\newpage
\begin{figure}[h]
\begin{center}
\epsfig{figure=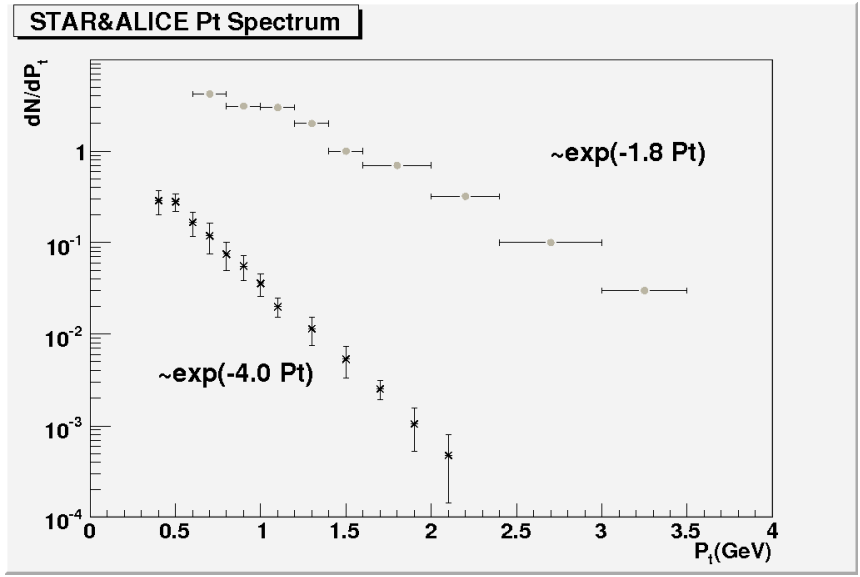,height=3.0in}
\caption{Transverse momentum distributions from STAR (200 GeV) and ALICE (900GeV), as fitted with the exponent.}
\end{center}
\end{figure}

\newpage
\begin{figure}[h]
\begin{center}
\epsfig{figure=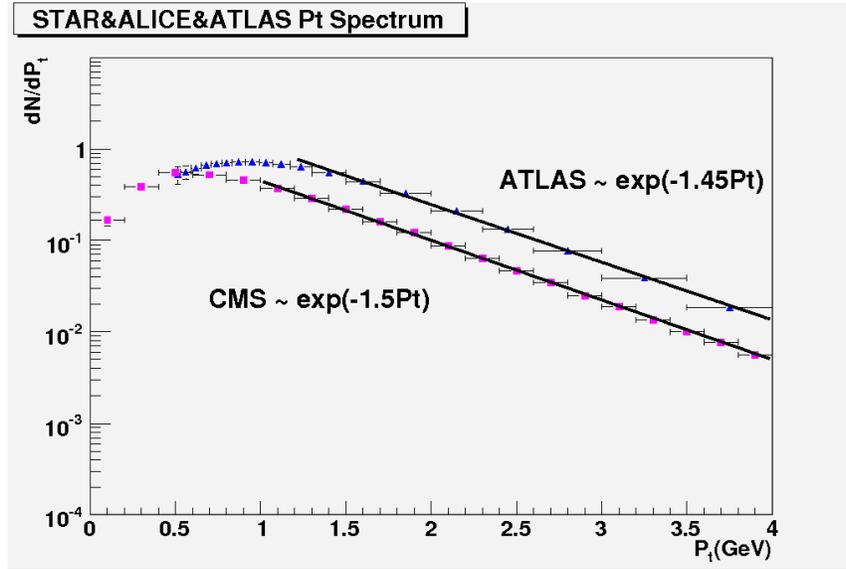,height=3.0in}
\caption{Transverse momentum distributions at $\sqrt{s}$= 7 TeV.}
\end{center}
\end{figure}

\newpage
\begin{figure}[h]
\begin{center}
\epsfig{figure=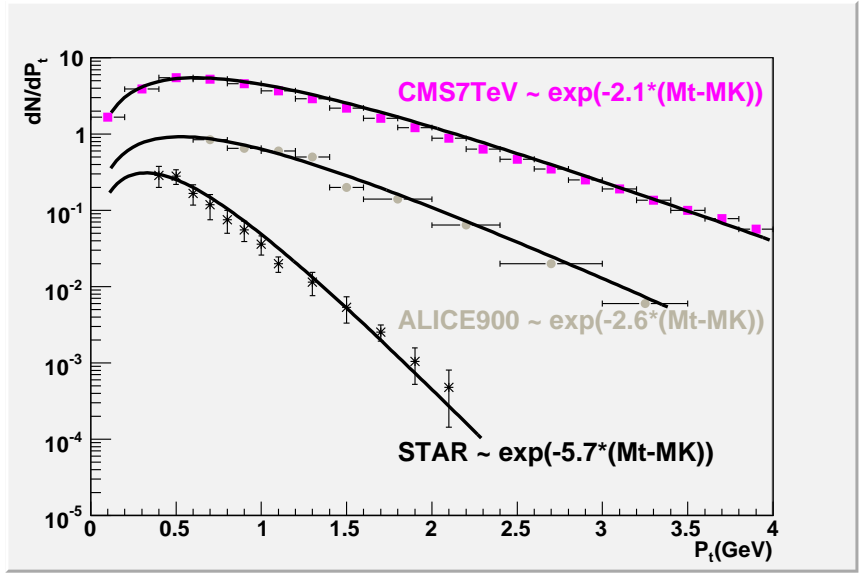,height=3.0in}
\caption{The description of STAR, ALICE and CMS data with the QGSM function.}
\end{center}
\end{figure}

\newpage
\begin{figure}[h]
\begin{center}
\epsfig{figure=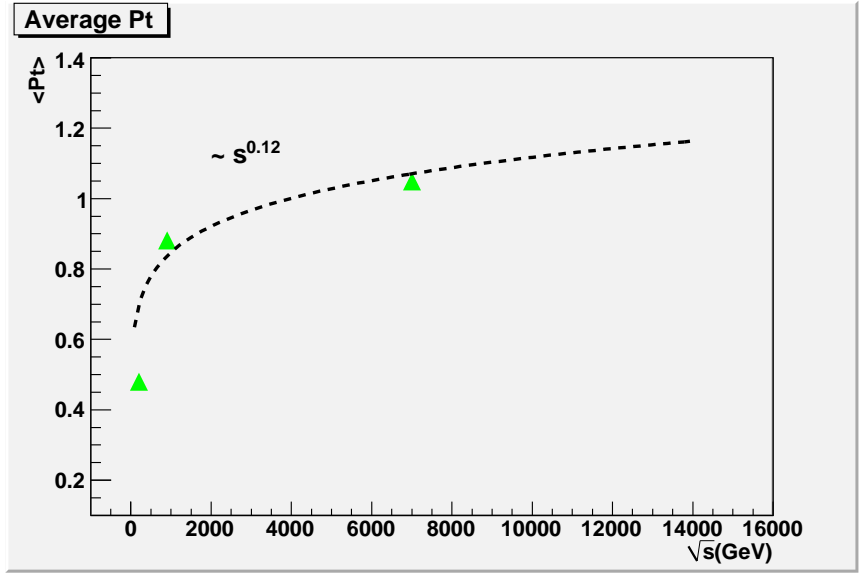,height=3.0in}
\caption{The average $p_t$ dependence on the energy and typical QGSM asympthotics $ ~ s^{0.12}$.}
\end{center}
\end{figure}

\end{document}